\documentclass[%
 reprint,
superscriptaddress,
 amsmath,amssymb,
 aps,
prb,
]{revtex4-2}
\usepackage{graphicx}% Include figure files
\usepackage{dcolumn}% Align table columns on decimal point
\usepackage{bm}% bold math
\usepackage{hyperref}% add hypertext capabilities
\hypersetup{colorlinks= True,
linkcolor=blue,
citecolor=blue,
urlcolor=blue}
\usepackage{xcolor}
\begin{document}

\preprint{APS/123-QED}

\title{Magnetic phase diagram and cluster glass-like properties of stage-1 graphite intercalated FeCl\texorpdfstring{$_3$}{3}}

\author{J.J.B. Levinsky}
\email{j.levinsky@rug.nl}
\affiliation{%
 Zernike Institute for Advanced Materials, University of Groningen, Nijenborgh 4, 9747 AG Groningen, The Netherlands
}%
\author{R. Scholtens}%
\affiliation{%
 Zernike Institute for Advanced Materials, University of Groningen, Nijenborgh 4, 9747 AG Groningen, The Netherlands
}%
\author{C. Pappas}
\affiliation{
 Faculty of Applied Sciences, Delft University of Technology, Mekelweg 15, 2629 JB Delft, The Netherlands
}%
\author{G.R. Blake}
\affiliation{%
 Zernike Institute for Advanced Materials, University of Groningen, Nijenborgh 4, 9747 AG Groningen, The Netherlands
}%

\date{\today}

\begin{abstract}
We present a comprehensive investigation of the magnetic properties of stage-1 graphite intercalated FeCl$_3$ using a combination of DC and AC magnetic susceptibility, thermoremanent magnetization and field dependent magnetization measurements. This van der Waals system, with a centrosymmetric honeycomb lattice, combines frustration and disorder, due to  intercalation, and may be hosting topologically non-trivial magnetic phases. Our study identifies two magnetic phase transitions at $T_{f1} \approx $ 4.2~K and at $T_{f2} \approx $ 2.7~K.
We find that the paramagnetic state, for $T>T_{f1}$, is dominated by short-range ferromagnetic correlations. These  build up well above $T_{f1}$ and lead to a significant change in magnetic entropy, which reaches $\Delta S_M^{Pk} =$ -5.52~J~kg$^{-1}$~K$^{-1}$ at 7~T. Between $T_{f1}$ and $T_{f2}$  we observe slow spin dynamics characteristic of a cluster glass-like state, whereas for $T$ $<$ $T_{f2}$ our results indicate the onset of a low temperature long range ordered state. The analysis of the experimental results leads to a complex phase diagram, which may serve as a reference for future investigations searching for topological non-trivial phases in this system.
\end{abstract}

\maketitle

\section{\label{sec:1}Introduction}

Frustrated centrosymmetric helimagnetic materials with trigonal symmetry are predicted to host exotic topological magnetic structures known as magnetic skyrmions \cite{Leonov2015, Lin2018, Okubo2012}. These swirling nanoscale spin vortices possess an inherent topological protection and are easily manipulated and moved by low current densities making them promising candidates for future spintronics applications \cite{Seki2016}. In this respect, van der Waals materials have had a considerable resurgence since the discovery of Ising-like out-of-plane ferromagnetism down to the monolayer limit in CrI$_3$ \cite{Huang2017a}. 

The van der Waals transition metal trihalide family of compounds exhibits a variety of interesting magnetic structures \cite{McGuire2017}. Their centrosymmetric honeycomb or triangular lattices give rise to frustrated magnetic exchange interactions, resulting in helimagnetic ordering in pristine FeCl$_3$ \cite{Cable1962}.
When the FeCl$_3$ layers are intercalated with graphite a whole family of two-dimensional (2D) helimagnetic materials can be realized, the FeCl$_3$-graphite intercalated compounds (GICs). In GICs the distance between the magnetic intercalant layers can be tuned by the manner in which the intercalant (in this case FeCl$_3$) is distributed throughout the graphite matrix according to a fixed periodicity. GICs are characterized by a stage index $n$, which describes the number of graphite layers between two adjacent intercalant layers. As the magnetic interlayer exchange coupling diminishes with increasing distance between the magnetic layers, the staging phenomenon provides a  mechanism for investigating the crossover from 3D to 2D magnetic behavior. For this reason, GICs have been considered model systems for the study of 2D magnetism \cite{Suzuki1998, Ibrahim1987}.  

The magnetic behavior of FeCl$_3$-GICs is not yet completely understood despite a considerable amount of research that has been carried out on these systems. Previous studies found that two main types of FeCl$_3$-GICs can be distinguished \cite{Rosenman1985}. The $\alpha$-type exhibits a single, stage-independent antiferromagnetic phase transition at 1.7~K, as determined by magnetic susceptibility and heat capacity measurements \cite{Ibrahim1987, Onn1982,Rosenman1985}. This transition has only been observed when a magnetic field is applied in the basal plane and is suppressed when $\mu_0H$ $\gtrsim$~1~mT \cite{Ibrahim1987}. The $\beta$-type GICs exhibit an antiferromagnetic phase transition at temperatures varying between 3.6~K and 5.5~K for stage-1 compounds \cite{Rosenman1985,Simon1983}. As the intercalant (FeCl$_3$) is extremely hygroscopic, it has been hypothesized that the intercalated $\alpha$-type can react with water in the air, creating the $\beta$-type variant \cite{Rosenman1985}. No structural changes have been identified in this transformation but a reduction of the ratio of Fe$^{3+}$ to Fe$^{2+}$ has been observed by M\"{o}ssbauer spectroscopy \cite{Rosenman1985}.

In stage-1 FeCl$_3$-GIC, this chemical reduction appears to involve charge transfer from the graphite host matrix to the FeCl$_3$ intercalant, as evidenced by the lowering of the Fermi energy \cite{Mele1979}. The acceptor site for the donated charge from the graphite has been the subject of debate for many years. Chlorine ions adjacent to iron vacancies at the periphery of intercalant islands were initially believed to be the sole acceptors \cite{Wertheim1980,Wertheim1981}. However, M\"{o}ssbauer spectroscopy reveals a more complex situation. Three distinct Fe-sites (A, B and C) are found in temperature dependent M\"{o}ssbauer measurements \cite{Millman1983,Millman1982a}. The majority component originates from site A and is attributed to Fe$^{3+}$ ions \cite{Millman1983}, the magnetic moments of which lie in the basal plane \cite{Millman1983,Millman1982a}. This component plays a dominant role in the onset of the low temperature magnetically ordered state, as only the Fe$^{3+}$ ions of site A were found to participate in the magnetic ordering \cite{Millman1982a}.

The minority Fe-sites are referred to as B and C. Site B is occupied by Fe$^{2+}$ ions, the easy axis of which is along the stacking direction \cite{Millman1983}, whereas site C is occupied by Fe$^{3+}$ ions that are directly adjacent to iron vacancies \cite{Millman1982b}. The respective concentrations of these two components are sample dependent but their  sum amounts to approximately 25\% of the total Fe sites for all samples investigated \cite{Millman1983}. We also note that the relative weight of the B-site component appears to be temperature dependent \cite{Millman1983,Millman1982a,Millman1982b}. Its contribution increases continuously upon cooling below 100~K \cite{Millman1983} reaching a sample dependent maximum concentration of approximately 17\% among all iron sites at 10~K. These results lead to a picture where, once all acceptor sites consisting of chlorine ions surrounding iron vacancies are exhausted, Fe$^{3+}$ ions act also as acceptor sites, yielding Fe$^{2+}$ at low temperatures \cite{Millman1983,Herber1984}. 

The transformation from the $\alpha$-phase to $\beta$-phase is almost unavoidable in powder samples \cite{Rosenman1985}. On the other hand, it is difficult to induce this transformation in samples based on highly oriented pyrolytic graphite (HOPG) or single-crystal Kish graphite (SCKG) \cite{Rosenman1985}. 

Powder neutron diffraction (PND) experiments showed that the stage-1 $\alpha$-type compounds order magnetically at 1.7~K in a configuration that can be described by an in-plane modulation vector of \textbf{Q} = 0.25~\textbf{a$^*$} and \textbf{Q} $\approx$~0.3~{\AA}$^{-1}$ \cite{Rosenman1985}, where \textbf{a}$^*$ is the reciprocal in-plane parameter of the iron lattice. On the other hand, the stage-1 $\beta$-type compounds undergo a magnetic phase transition at 3.8~K to a low temperature 3D long range ordered phase characterized by an in-plane incommensurate modulation with a vector \textbf{Q} = $0.394 \; \textbf{a}^*$ and \textbf{Q} $\approx$~0.467~{\AA}$^{-1}$ \cite{Rosenman1985,Simon1983}.

The analysis of the magnetic diffraction peak shapes 
shows that this transition corresponds to a crossover from a 2D to 3D ordering \cite{Rosenman1985,Simon1983}. Indeed, whereas  the  peak shapes are symmetric below 3.8~K, as expected for 3D long range ordering, they become asymmetric above 3.8~K   adopting  Warren-type   shapes characteristic of powder patterns of 2D systems.

Warren-type asymmetric peak shapes have been reported at all temperatures for the $\beta$-type stage-2 compounds \cite{Simon1983}. Thus, in this system the magnetic correlations are strictly 2D in nature although  they can be described using the same modulation vector as for the stage-1 compound \cite{Simon1983}. Both stage-2 and stage-3 compounds show spin glass-like behavior \cite{Suzuki1998}. For the stage-3 compound AC magnetic susceptibility hints at a spin-glass state emerging at a characteristic temperature that increases with increasing frequency, $f$, and is equal to 3.7~K for $f$ = 3.7 Hz \cite{Miyoshi1997}. For the stage-2 compound, two separate spin-glass-like transitions have been reported at $T_h \approx$~4.5 - 6.1~K and $T_l \approx$~2 - 2.5~K, which also shift to higher temperatures with increasing frequency \cite{Suzuki1998}.

Reviewing the known literature, it becomes clear that discrepancies exist between the magnetic behavior observed via magnetometry, neutron scattering and M\"{o}ssbauer spectroscopy. The nature of the magnetic phase transition observed in stage-1 FeCl$_3$-GIC, its exact ordering temperature and the evolution of the magnetic phase under  applied magnetic fields are still largely unknown. As the magnetic properties of stage-2 FeCl$_3$-GIC have been studied in more detail and are better understood than stage-1 FeCl$_3$-GIC, there exists a gap in knowledge on the evolution of the magnetic properties of the FeCl$_3$ system as it undergoes intercalation and the interlayer exchange is diminished.

In the following we discuss the magnetic properties and phase diagram of polycrystalline stage-1 FeCl$_3$-GIC based on the analysis of temperature and frequency-dependent AC and DC magnetic susceptibility, field-dependent magnetization and time-dependent thermoremanent magnetization measurements. We conclude that, at zero magnetic field and for $T_{f2} \lesssim $~2.7~K the ground state is a long range ordered state, which is preceded by a cluster glass-like phase that sets-in for $T_{f1} \lesssim $~4.2~K. Furthermore, in the paramagnetic phase, short-range ferromagnetic correlations build up with decreasing temperature, which lead to a significant change in the magnetic entropy. Our results lead to a magnetic phase diagram that accounts for all magnetic phase transitions, as a function of temperature and magnetic field, and which can serve as a reference for future investigations searching for topological non-trivial phases in this system.

\begin{figure}[t]
    \centering
    \includegraphics[width = 8.6 cm]{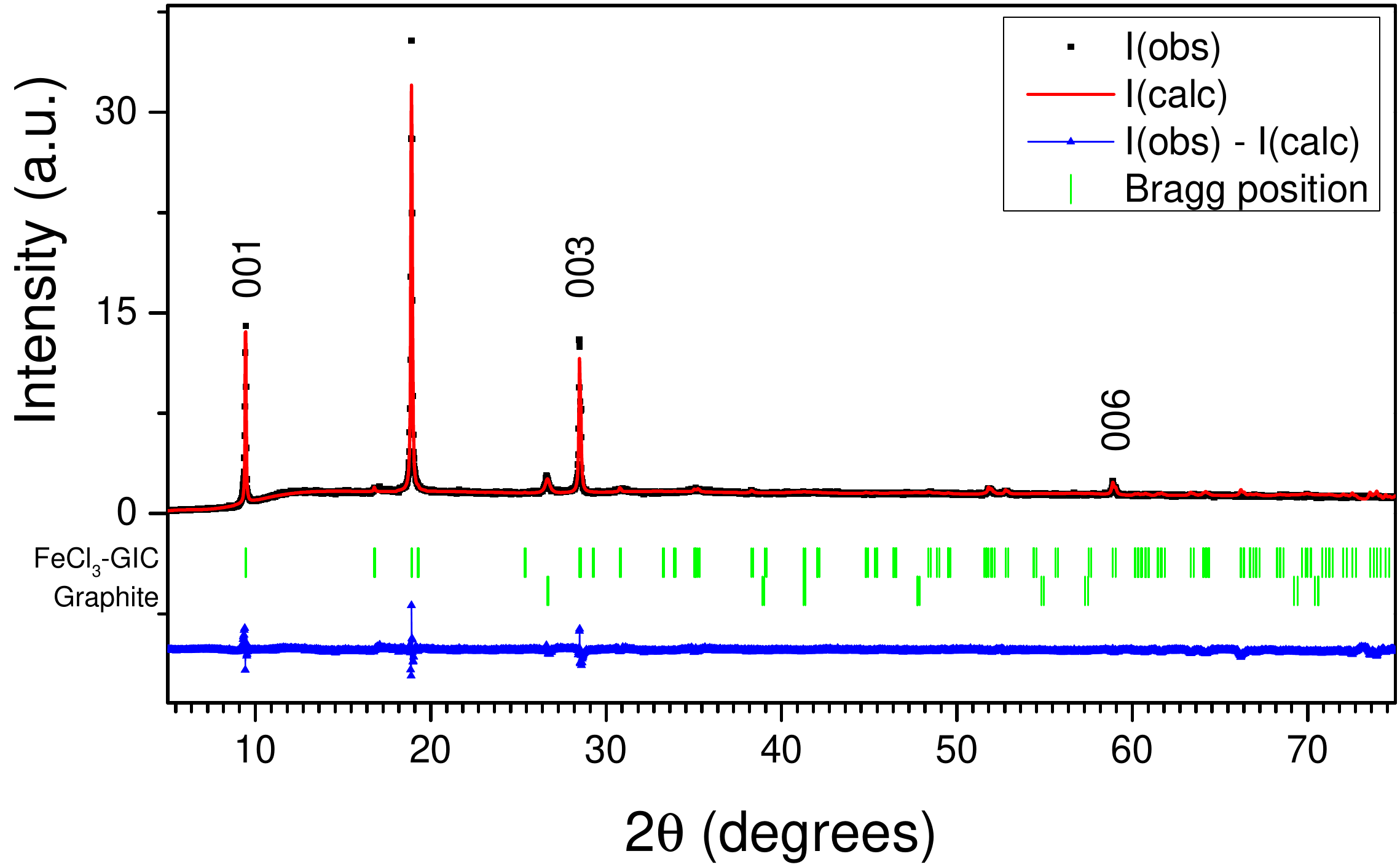}
    \caption{Refined powder XRD data of our stage-1 FeCl$_3$-GIC sample with selected $00 \ell$ reflections indicated. The green markers indicate the expected peak positions for stage-1 FeCl$_3$-GIC (upper) and graphite (lower). The difference between I(calc) and I(obs) is plotted below the phase markers.}
    \label{fig:1}
\end{figure}

\section{\label{sec:2}Experimental methods}
The single-phase, polycrystalline samples of stage-1 FeCl$_3$-GIC used for this work were synthesized by a single tube method. A 4:1 mass ratio of anhydrous FeCl$_3$ to graphite powder ($<$ 150 $\mu$m) was added to a Pyrex tube which was sealed under rough vacuum ($P$ $\approx$~10$^{-3}$ bar). The tube was placed in a tube furnace and held at 300~$^{\circ}$C for 60 hours. Consequently, the empty side of the tube was cooled down 2 hours earlier than the side containing the graphite and FeCl$_3$, which was cooled down to room temperature over the course of 10 hours. This was done to ensure that excess FeCl$_3$ does not condense on the surface of the intercalated product. The product was then washed with 2 wt\% HCl solution and deionized water. The product was stored in a nitrogen-filled glove box to prevent further degradation by exposure to the air.
The quality of the sample was checked by  X-ray diffraction on a Bruker D8 Advance powder diffractometer operating in Bragg–Brentano geometry with Cu K$\alpha$ radiation ($\lambda$ = 1.5406 \AA).

The DC magnetization, $M$, was measured as a function of temperature, $T$, and the applied magnetic field, $\mu_0 H$, using a Quantum Design MPMS XL-7 T SQUID magnetometer. The real and imaginary components of the AC magnetic susceptibility, $\chi^\prime$ and $\chi^{\prime\prime}$ respectively, were determined using the ACMS II option for the Quantum Design PPMS system. The data were corrected for the diamagnetic contribution of graphite by subtracting an experimentally determined temperature independent diamagnetic susceptibility of -2.97$\times$10$^{-8}$ m$^3$/mol.

\section{\label{sec:3}Experimental results}

\subsection{\label{sec:3a}Powder X-ray diffraction}

A characteristic powder X-ray diffraction (PXRD) pattern of our polycrystalline FeCl$_3$-GIC samples is shown in Fig. \hyperref[fig:1]{1} together with the fit obtained using the Le Bail method in the Jana2006 software package \cite{Petricek2014}. Graphite intercalation compounds can be identified and checked for stage uniformity by inspecting the $00 \ell$ diffraction peaks \cite{Suzuki1998, Miyoshi1997, Dresselhaus2002}. Our sample shows a single set of $00 \ell$ peaks with a c-axis lattice parameter of 9.4060(3)~\AA. This corresponds to pure stage-1 FeCl$_3$-GIC and as expected for FeCl$_3$-GICs, the 002 reflection of free graphite is also observed \cite{Cowley1956}. 

\begin{figure}[t]
    \centering
    \includegraphics[width = 8.6 cm]{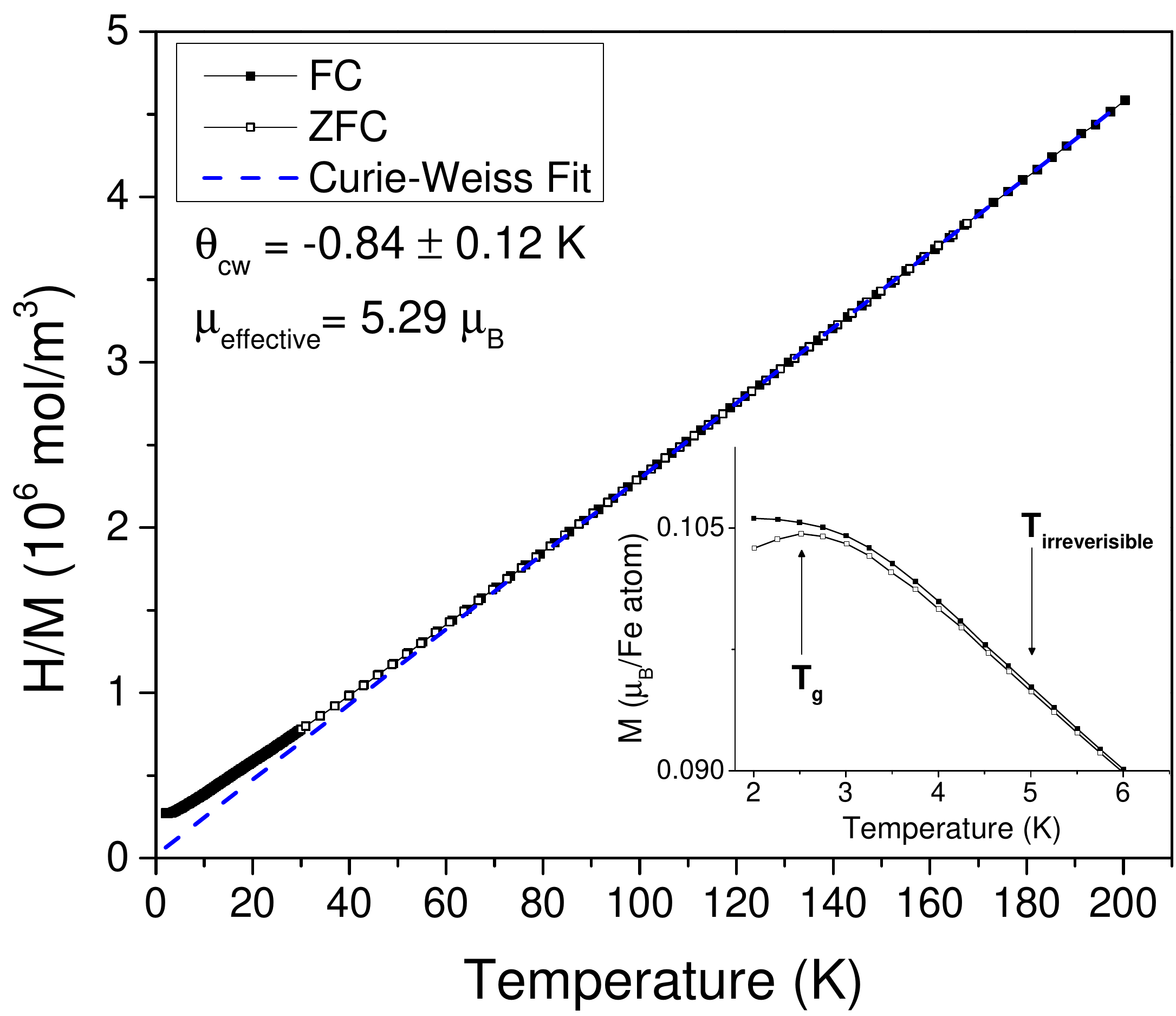}
    \caption{Field-cooled (FC) and zero-field-cooled (ZFC) inverse field normalised magnetization, H/M, plotted versus temperature. The dashed line indicates the Curie-Weiss law fit to the data for $T>$ 100~K, leading to the parameters shown in the figure and listed in Table \ref{tab:table1}. The inset shows a close-up of the low temperature region (2-9 K) to highlight the observed peak at $T_g$(200~mT) = 2.5~K and the difference between the field cooled (FC) and zero field cooled (ZFC) curves.}
    \label{fig:2}
\end{figure}
\begin{figure*}[t!]
    \centering
    \includegraphics[width = 15.5 cm]{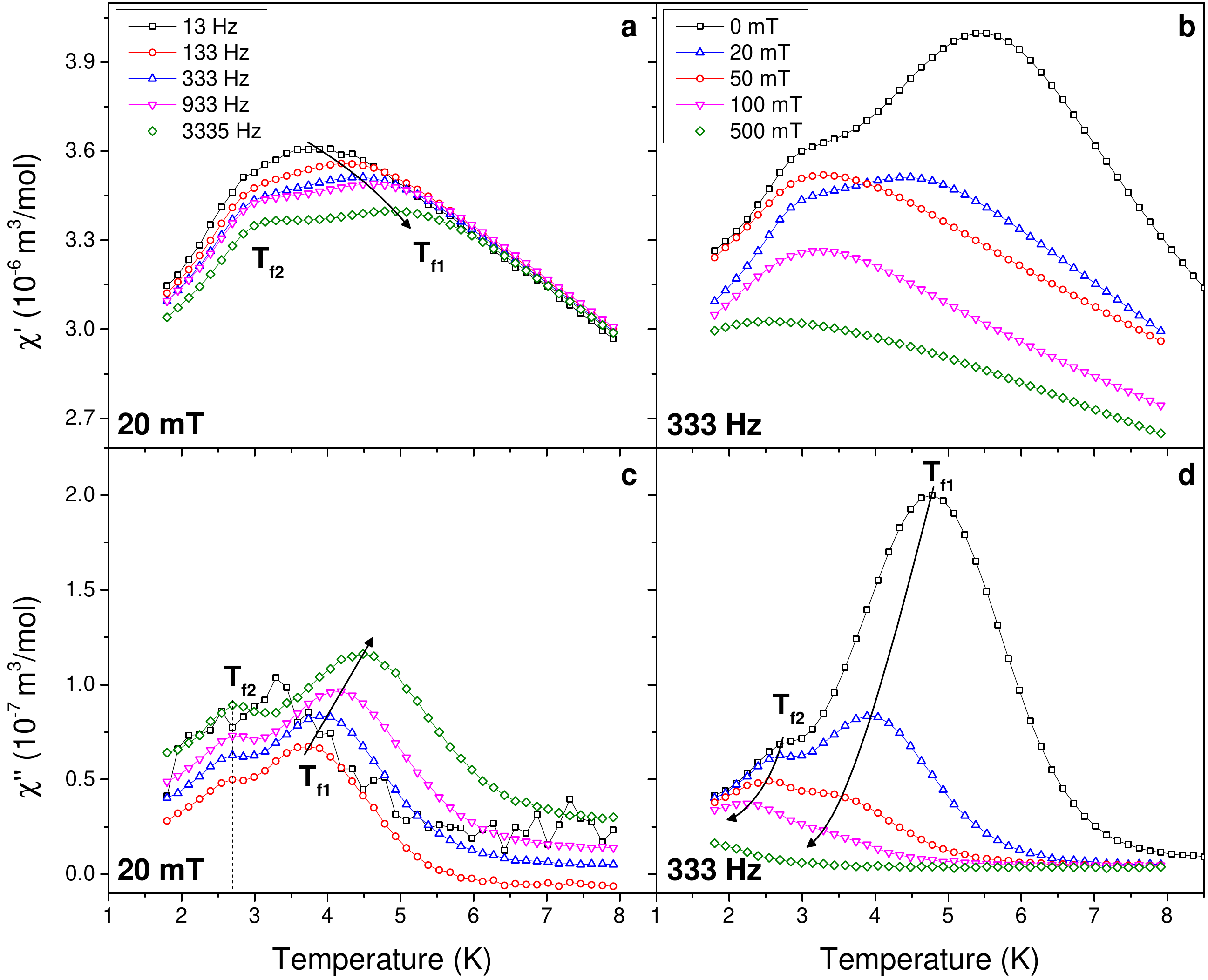}
    \caption{Real (a and b) and imaginary (c and d) components of the AC susceptibility plotted against temperature. In (a) and (c) the measurements were performed under a DC bias of 20~mT with a superimposed AC signal of 0.43~mT oscillating at the five different frequencies indicated in the legends. In (b) and (d) the measurements were performed under DC applied magnetic fields of which the strength is indicated in the legends and with a superimposed AC signal of 0.43~mT oscillating at 333~Hz. Note that the data for a DC field of 20~mT and a frequency of 333~Hz are depicted in both (a), (b) and (c), (d) (blue triangles). }
    \label{fig:3}
\end{figure*}

\begin{table*}[hbt!]
\caption{\label{tab:table1}%
Comparison of the Curie-Weiss parameters  determined by  previous studies and in the current work (shown on the last line and indicated by an asterisk).
}
\begin{ruledtabular}
\begin{tabular}{c c c}
\textrm{Fitting range}&
\textrm{Curie-Weiss temperature (K)}&
\textrm{$\mu_{eff}$ ($\mu_B$ per Fe atom)}\\
\colrule
100 – 300~K \cite{Hohlwein1974} & 10 ($\parallel$ c-axis) / 1 ($\perp$ c-axis) & 5.49 ($\parallel$ c-axis) / 5.56 ($\perp$ c-axis)\\
40 – 100~K \cite{Hohlwein1974} & 0 ($\parallel$ c-axis) / -5 ($\perp$ c-axis) & -\\
25 – 77~K \cite{Ohhashi1974} & -11.4 ($\parallel$ c-axis) / -8.2 ($\perp$ c-axis) & 5.87 ($\parallel$ c-axis) / 5.98 ($\perp$ c-axis) \\
20 – 60~K \cite{Millman1982} & 3.8 ($\parallel$ c-axis) / -3.8 ($\perp$ c-axis) & - \\
100 – 200 K* & -0.8 (powder) & 5.29
\end{tabular}
\end{ruledtabular}
\end{table*}

\subsection{\label{sec:3b}DC magnetization at 200~mT}

Fig. \hyperref[fig:2]{2} shows the temperature dependence of $H/M$, where $M$ is the magnetization measured under an applied magnetic field of $\mu_0 H$ = 200~mT. At high temperatures, for $T >$ 80~K, $H/M$ increases linearly with increasing $T$ as expected from  the  Curie-Weiss  law. A linear fit yields an effective magnetic moment of $\mu_{eff}$ = 5.29~$\mu_B$ and a Curie-Weiss temperature of $\theta_{cw}$ = -0.8~K.

Below 80~K the inverse susceptibility starts to deviate from the Curie-Weiss behavior in a continuous fashion indicating an increase of both the effective magnetic moment and the antiferromagnetic character of the interactions with decreasing temperature. These observations indicate the onset of short-range order and are consistent with PND results \cite{Simon1983}, according to which short-range magnetic correlations gradually build up below 30~K in this system. 
Table \hyperref[tab:table1]{I} compares our Curie-Weiss law parameters with those reported in the literature. The  scatter in the Curie-Weiss temperatures reveals a significant sample dependency of the magnetic properties of stage-1 FeCl$_3$-GIC, for reasons that we discuss below.

\begin{figure*}
    \centering
    \includegraphics[width = 17.2 cm]{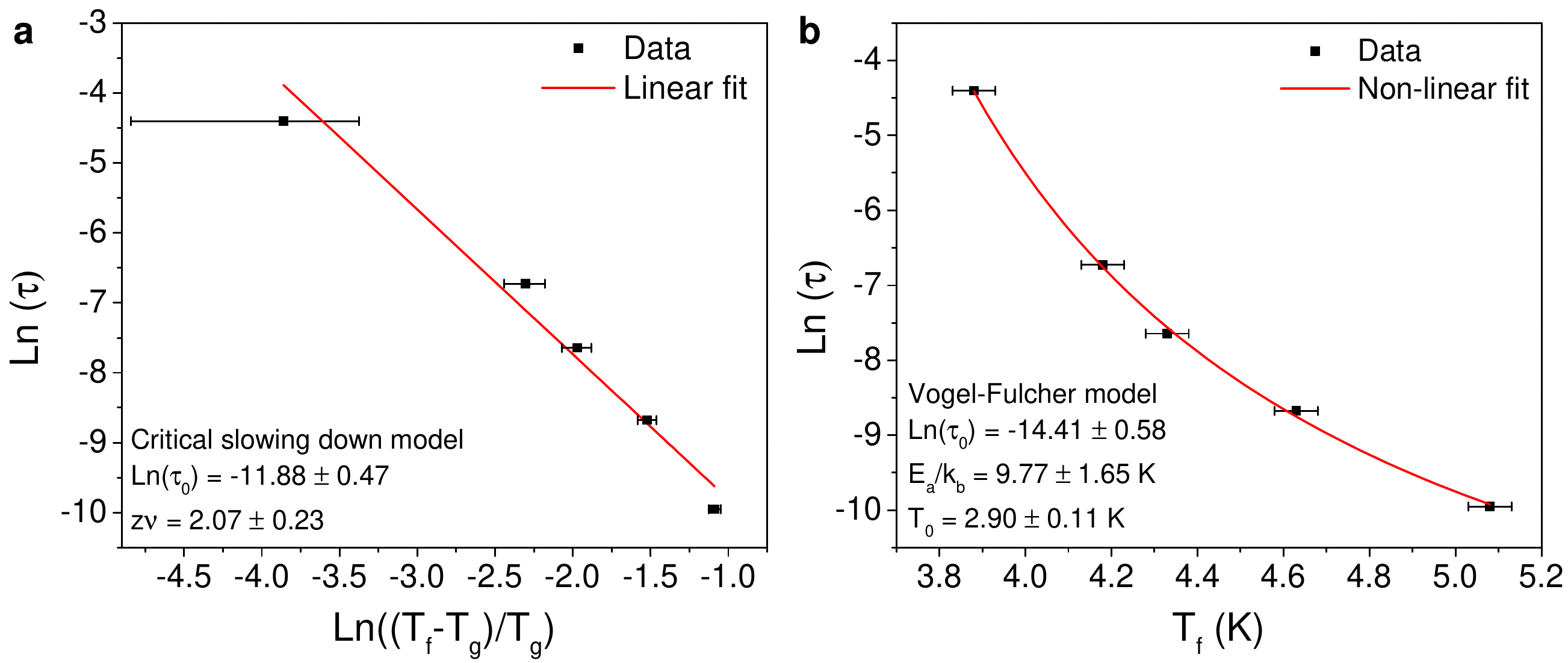}
    \caption{Influence of the characteristic time corresponding to the frequency of the measurement, $\tau$ = 1/$\omega$ = 1/$2 \pi f$, on the freezing temperature $T_f$. The same data are fitted against (a) the critical slowing down model and (b) the Vogel-Fulcher model. The error bars represent the uncertainty (0.05~K) in the determination of the freezing temperature.}
    \label{fig:4}
\end{figure*}
    
Our effective magnetic moment is smaller than the theoretical value of 5.94~$\mu_B$. A similar deviation from the theoretical effective moment has also been reported by Hohlwein et al. \cite{Hohlwein1974} and is most likely due to a deviation of the actual molecular mass, $\mathbb{M}$,  from its ideal value. When  fully intercalated, stage-1 FeCl$_3$  has the composition C$_{6.2}$FeCl$_3$ which leads to $\mathbb{M}$ = 0.23666~kg/mol. However, as the degree of intercalation appears to be sample  as well as time dependent, the actual composition of the sample may deviate from the ideal composition,  affecting the value of the deduced effective magnetic moment.

The degree of intercalation should also influence the number of iron sites that are next to vacancies. This would change the amount of Fe$^{2+}$ present in the sample and lead to sample-dependent magnetic properties. Furthermore, subtraction of the significant diamagnetic contribution of graphite from the magnetic susceptibility affects the determination of both the Curie-Weiss temperature and the effective moment. Thus far, this only seems to have been taken into account in the analysis of the magnetization measurements performed by Ohhasi and Tsujikawa \cite{Ohhashi1974}.

The inset of Fig. \hyperref[fig:2]{2} shows the  temperature dependence of the zero-field-cooled (ZFC) and field-cooled (FC) DC magnetization for $T$ $<$ 7~K,  measured with increasing temperature under an applied magnetic field of $\mu_0 H$ = 200~mT.  The ZFC and FC curves split below $T_{irreversible} \approx$~5~K, and show a behavior commonly seen for spin glasses, cluster glasses, superparamagnetic and spin liquid systems \cite{Kulish2019}. However, the ZFC curve shows a well-defined peak at a much lower temperature, $T_g$(200~mT) = 2.5~K. As we will discuss below, when the field is decreased to 3~mT this peak shifts to a higher temperature, $T_g$(3~mT) = 3.8~K (see Fig. S1 in the Supplementary Material \cite{SM}), and thus shifts towards $T_{irreversible}$ with decreasing magnetic field. 

\subsection{\label{sec:3c}AC susceptibility}

In order to investigate the nature of the magnetic phase transitions at $T_{irreversible}$ and $T_g$ we performed AC susceptibility measurements in zero-field-cooled conditions for five values of DC bias ($\mu_0 H$ = 0, 20, 50, 100 and 500~mT) with a superimposed AC signal of $\mu_0 H$ = 0.43~mT oscillating at five different frequencies: $f$ = 13, 133, 333, 933 and 3335 Hz. Fig. \hyperref[fig:3]{3(a)} and \hyperref[fig:3]{3(c)} show $\chi^\prime$ and $\chi^{\prime\prime}$, measured under a DC bias of 20~mT and for all five frequencies, plotted against temperature  (see Fig. S2 in the Supplementary Material \cite{SM}, for plots of  $\chi^\prime$ and $\chi^{\prime\prime}$ obtained under the other mentioned DC biases).  Complementary information is provided by Fig. \hyperref[fig:3]{3(b)} and \hyperref[fig:3]{3(d)}, which depicts the temperature dependence of $\chi^\prime$ and $\chi^{\prime\prime}$ measured for $f$ = 333~Hz and for all five magnetic field strengths.

The $\chi^{\prime}(T)$ curves in Fig. \hyperref[fig:3]{3(a)}, which account for the reversible magnetic response of the system \cite{Kulish2019,Baanda2013}, reveal a maximum at $T_{f1}$ = 4.2~K and a shoulder at $T_{f2}$ = 3~K for $f$ = 133~Hz. With increasing frequency, the maximum at $T_{f1}$ shifts to higher temperatures while the shoulder at $T_{f2}$ remains unchanged within the resolution of our measurements. The magnitude of both maxima decreases monotonically with increasing frequency. $T_{f1}$ and $T_{f2}$ are correlated to $T_{irreversible}$ and $T_g$ observed in the DC magnetization measurements (section \hyperref[sec:3b]{III.B}), respectively. 

The effect of the magnetic field on $\chi^{\prime}(T)$, for $f$ = 333~Hz, is shown in Fig. \hyperref[fig:3]{3(b)}. With increasing field, both observed maxima decrease in magnitude. However, the magnetic field does not affect the two maxima in the same way. The maximum at $T_{f1}$  shifts markedly towards lower temperatures with increasing magnetic field and its intensity decreases so dramatically that it is no longer observable for $\mu_0 H >$ 20~mT. On the other hand, the position of the peak at $T_{f2}$ remains almost unchanged for $\mu_0 H \leq$ 20~mT before shifting to lower temperatures with increasing magnetic field.

The $\chi^{\prime\prime}(T)$ curves in Fig. \hyperref[fig:3]{3(c)} and \hyperref[fig:3]{3(d)}, which account for the irreversible magnetic response of the system \cite{Kulish2019,Baanda2013}, bear a clear signature of both transitions and show maxima at $T_{f1}$ = 3.7~K and $T_{f2}$ = 2.7~K for $f$ = 133~Hz. At $\mu_0 H$ = 20~mT the maximum at $T_{f1}$ shifts to higher temperatures with increasing frequency, while the position of the peak at $T_{f2}$ does not change significantly. The effect of the magnetic field  on $\chi^{\prime\prime}(T)$ for $f$ = 333~Hz is shown in Fig. \hyperref[fig:3]{3(d)}. With increasing magnetic field, both peaks decrease in magnitude and shift towards lower temperatures. Similarly to the behavior found for $\chi^{\prime}(T)$, the peak at $T_{f1}$ becomes too weak to be observed for ${\mu}_0H \geq$ 100~mT. Above this field the main contribution to  $\chi^{\prime\prime}(T)$ is related to the transition at $T_{f2}$.

The frequency-dependent maxima in $\chi^{\prime}(T)$ bear similarities with glassy magnetic systems \cite{Mydosh1993}. To study and classify this behavior, we determined the values of $T_{f1}$ for every frequency and used them to extract the Mydosh parameter $\delta$ \cite{Mydosh1993}:

\begin{equation}
    \delta = \frac{\Delta T_f}{T_f \; \Delta (\log _{10}.  \omega)}.
    \label{eq:1}
\end{equation}

\noindent Here, $T_f$ is the freezing temperature at zero frequency and $\Delta T_f$ is the difference in the freezing temperatures for a corresponding difference of the angular frequencies $\omega$ = $2\pi f$. The Mydosh parameter thus quantifies the relative shift of the freezing temperature per  frequency decade. 

Our data lead to $\delta$ = 0.13, which is much lower than $\delta$ = 0.28, the value reported for non-interacting ideal superparamagnetic systems such as $\alpha$-(Ho$_2$O$_3$)(B$_2$O$_3$) \cite{Mydosh1993}. On the other hand, our value of $\delta$ is higher than the values reported for canonical spin-glasses (0.005 $<$ $\delta$ $<$ 0.06) \cite{Mydosh1993} or cluster glasses ($\delta$ in the range 0.01 – 0.09) \cite{Kulish2019,Malinowski2011a}.  The Mydosh parameter for stage-1 FeCl$_3$-GIC thus appears to take an intermediate value slightly larger than that of cluster glasses but significantly lower than that of superparamagnets.

As a next step we analyzed the frequency dependence of $T_{f1}$ defined as the maximum of the $\chi^{\prime}(T)$ curves under an applied field of $\mu_0H$ = 20~mT. The Arrhenius law, which corresponds to the simplest thermal activation model in the absence of any interactions, yields unphysical values of the parameters and will not be considered further. In the following we will analyze our data using one model based on dynamic scaling theory and the Vogel-Fulcher law.  

In the presence of a second order phase transition to a spin glass state at a temperature $T_g$, which ideally is determined at zero frequency \cite{Mydosh1993,Hohenberg1977,Binder1986, Pappas:1985hl,Pappas:2003cq}, the time dependence of $T_{f1}$ should follow the critical slowing down predicted by dynamic scaling theory \cite{Hohenberg1977}:

\begin{equation}
    \tau = \tau_0
    \left(\frac{T_f - T_g}{T_g}\right) ^{-z\nu},
    \label{eq:2}
\end{equation}

\noindent with $\tau = 1/f$, $\tau_0$ the characteristic relaxation time of the system, $z$  the dynamic critical exponent and $\nu$ the critical exponent for the correlation length. Equation \hyperref[eq:2]{2} can be rewritten as:

\begin{equation}
    \ln (\tau) = \ln (\tau_0) - z \nu  \; \ln 
    \left(\frac{T_f - T_g}{T_g}\right).
    \label{eq:3}
\end{equation}

In Fig. \hyperref[fig:4]{4(a)}, we plot $\ln (\tau)$ against $\ln((T_f-T_g)/T_g)$, with  $T_g$ = 3.8~K, as determined by DC magnetization measurements at 3~mT (see Fig. S1 in the Supplemental Material \cite{SM}). From the slope and the intercept of the linear fit, we obtain $z \nu$ = 2.07 $\pm$ 0.23 and $\tau_0$ = (7.7 $\pm$ 3.4) $\cdot$ 10$^{-6}$~s. This value of $z\nu$ is significantly smaller than that typically expected for spin-glass behavior, 4 $<$ $z\nu$ $<$ 12 \cite{Mydosh1993,Anand2012}.
Also the value of $\tau_0$ is much larger than for canonical spin-glasses, where it is of the order of 10$^{-12}$ s. This high value of $\tau_0$ reflects slow dynamics, indicating the presence of large correlated volumes.

The empirical Vogel-Fulcher (VF) law \cite{Mydosh1993,Vogel1921,Fulcher1925} has originally been proposed to describe the temperature dependence of the viscosity of supercooled liquids:

\begin{equation}
    \tau = \tau_0 \exp 
    \left(\frac{E_a}{k_B \; (T_f-T_0)}\right), 
    \label{eq:4}
\end{equation}

\noindent where $E_a$ is an activation energy and $T_0$ is the Vogel-Fulcher temperature, which  can be considered as a measure of the inter-cluster interaction strength \cite{Mydosh1993,Anand2012,Malinowski2011a}. Equation \hyperref[eq:4]{4} can be rewritten as:

\begin{equation}
    T_f = \frac{E_a}{k_B  \; \ln (\tau/\tau_0)} +T_0, 
    \label{eq:5}
\end{equation}

\noindent leading to the plot of  Fig. \hyperref[fig:4]{4(b)}. A fit of our data to the VF law without fixing any parameters,  leads to: $T_0$ = 2.9 $\pm$ 0.1~K, $\tau_0$~=~(6.5 $\pm$ 3.4) $\cdot$ 10$^{-7}$~s and $E_a/k_B$ = 9.8 $\pm$ 1.7~K. Thus, $\tau_0$ is comparable to the value found using the critical slowing down approach of equation \hyperref[eq:2]{2}. Furthermore, $E_a/k_B$ and $T_0$ are comparable to the ordering temperature, which validates the use of the VF law for this system.

Our analysis indicates that stage-1 FeCl$_3$-GIC undergoes a transition at $T_{f1}$ to a spin-glass-like phase characterized by relatively slow dynamics. The characteristic relaxation times ($\tau_0$) are of the order of 10$^{-7}$~s, indicating the presence of correlated clusters of spins instead of individual magnetic moments. We therefore conclude that the magnetic phase below $T_{f1}$ behaves like a cluster spin-glass phase. The fact that the Vogel-Fulcher temperature is of the same magnitude as the activation energy, $T_0 \propto E_a$⁄$k_B$, implies that the interactions between clusters are  of intermediate strength \cite{Malinowski2011a}.

The frequency and field dependent behavior of $T_{f1}$ shares significant similarities with that reported for stage-2 FeCl$_3$-GIC \cite{Suzuki1998}. A key difference between these two systems, however, is in the behavior of $T_{f2}$, which does not change with frequency. This is another indication that stage-1 FeCl$_3$-GIC undergoes a transition to a long-range 3D antiferromagnetic state at $T_{f1}$, in agreement with previous PND results \cite{Simon1983}.

\begin{figure}[b]
    \centering
    \includegraphics[width = 8.6 cm]{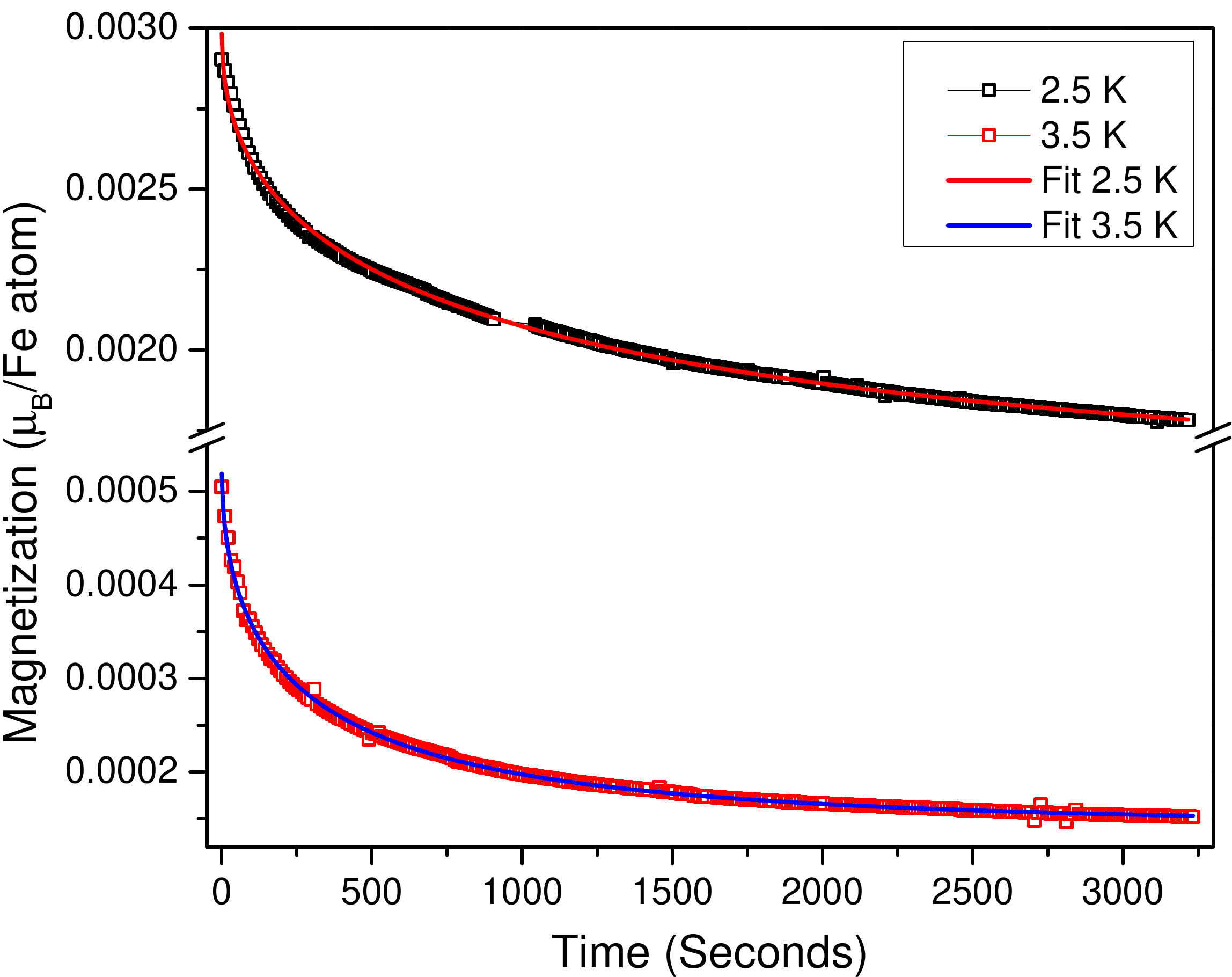}
    \caption{Thermoremanent magnetization versus time, measured after cooling in 1~T to either 3.5 or 2.5~K, and switching off the field. The solid lines indicate fits to the stretched exponential function of equation \hyperref[eq:6]{6}.}
    \label{fig:5}
\end{figure}

\subsection{\label{sec:3d}Thermoremanent magnetization}

In order to investigate the slow dynamics associated with the cluster glass-like phase we performed thermoremanent magnetization (TRM) measurements at 2.5~K and 3.5~K. The sample was cooled under an applied magnetic field of 1~T from 50~K to the target temperature after which the field was switched off and the magnetization was measured as a function of time. The resulting decay of the magnetization as a function of time is depicted in Fig. \hyperref[fig:5]{5}. 

The magnetization does not follow a simple logarithmic or power law decay. The best fit to the data was achieved using the stretched exponential function, which commonly describes the magnetization decay of spin- and cluster-glasses \cite{Mukadam2005}:

\begin{equation}
    M(t) = M_0 + M_r \exp \left[-\left(\frac{t}{\tau_0}\right)^{1-n}\right].
    \label{eq:6}
\end{equation}

\begin{table}[b]
\caption{\label{tab:table2}%
Parameters of the stretched exponential function fitting the decay of the magnetization at 2.5 and 3.5~K.\\}
\begin{ruledtabular}
\begin{tabular}{c|cc}
\textrm{}&
\textbf{2.5 K}&
\textbf{3.5 K}\\
\colrule
\bm{$M_0$} ($\mu_B/$Fe atom) & 0.00152(2) & 0.00014(0)\\
\bm{$M_r$} ($\mu_B/$Fe atom) &  0.00147(2) & 0.00038(2)\\
\bm{$\tau$} (s) & 1074(39) &  292(3)\\
\bm{$n$} & 0.517(8) & 0.464(5) 
\end{tabular}
\end{ruledtabular}
\end{table}

\noindent Here, $M_0$ is a time-independent ferromagnetic-like component, $M_r$ is the time-dependent (relaxing) magnetization and $\tau_0$ the characteristic relaxation time. The parameter $n$ is the stretching exponent \cite{Mukadam2005,Sato2016}: for $n$ = 0 the relaxation is exponential, reflecting a single time-constant, while for 0 $<$ $n$ $<$ 1 the relaxation is a stretched exponential, reflecting a distribution of relaxation times.

Table \ref{tab:table2} shows the values of the parameters derived from the fit of equation \hyperref[eq:6]{6} to the TRM data. When the temperature decreases from 3.5~K to 2.5~K, the relaxation time increases significantly, from 292 to 1074~s and $M_0$ increases by an order of magnitude. On the other hand, the ratio $M_r/M_0$ changes less dramatically, from approximately 2.6 to 1. Also the stretching exponent does not change and is equal to $n \approx$~0.5, a value similar to that found in structural glasses, indicating a broad distribution of relaxation times \cite{Sato2016}.

Our analysis shows that $M_r/M_{sat}$ = 0.03\% and 0.008\% at 2.5 and 3.5~K respectively, with similar ratios for $M_0/M_{sat}$, where $M_{sat} \approx 5 \; \mu_B$/Fe atom as expected for for Fe$^{3+}$ ions.  This is a very small fraction of the total magnetic moment and could arise from the Fe$^{2+}$ and Fe$^{3+}$ ions on the B and C sites respectively, which according to M\"{o}ssbauer spectroscopy, do not participate in the long range order \cite{Millman1982a}. However, some coupling might exist between  these magnetic moments, or clusters of magnetic moments, and the long range ordered phase. This may explain  the persistence of TRM below $T_{f2}$ as well as the long and temperature dependent relaxation times listed in Table \ref{tab:table2}. 

\subsection{\label{sec:3e}Field dependent magnetization}

\begin{figure}[t]
    \centering
    \includegraphics[width = 8.6 cm]{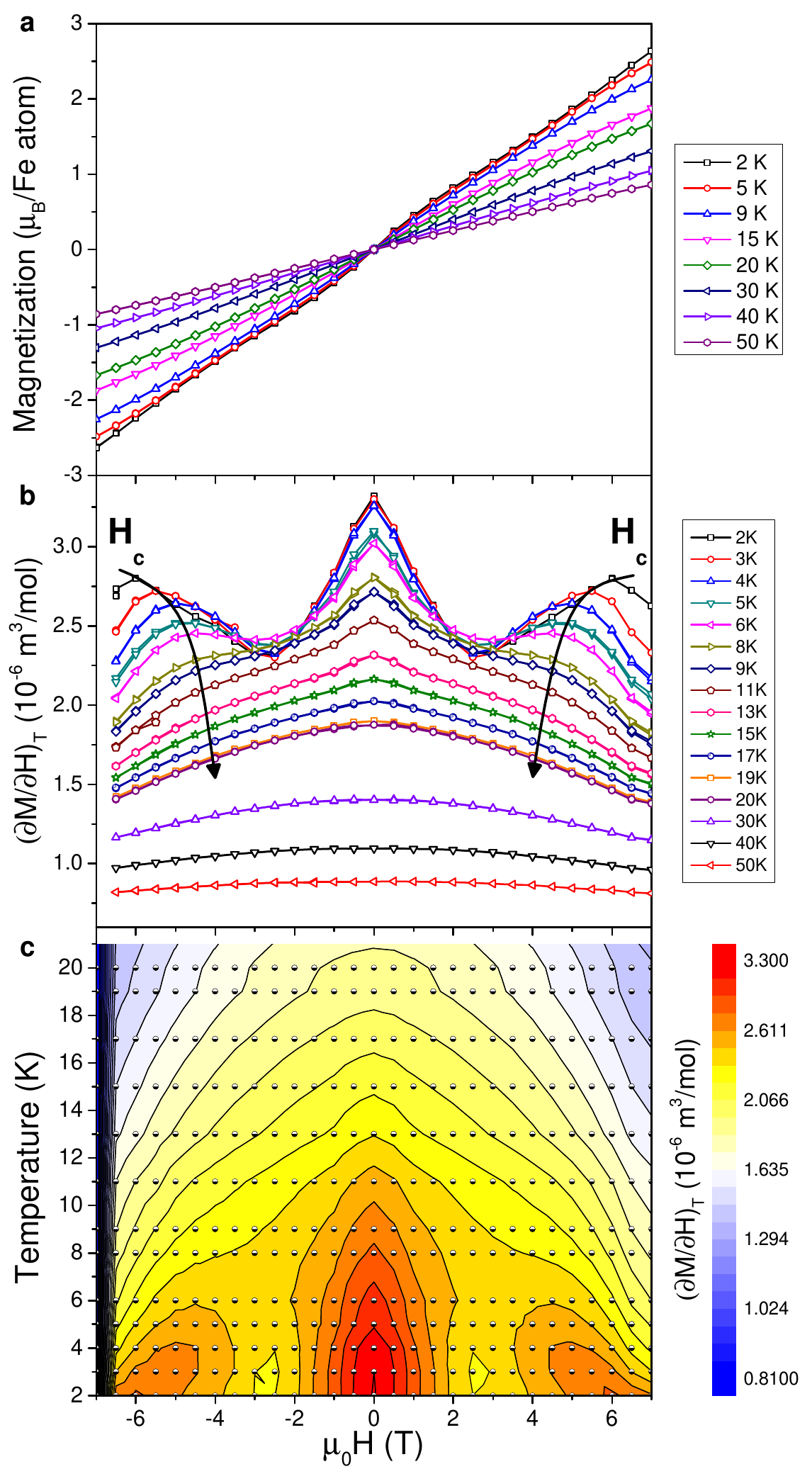}
    \caption{ Magnetic field dependence of (a) the magnetization and (b) the numerical derivative, $(\partial M / \partial H)_T$, versus applied magnetic field. The characteristic magnetic field $\mu_0 H_c$ seen in (b) is indicated by a black arrow. (c) Contour plot of $(\partial M / \partial H)_T$ versus temperature and magnetic field. The half open dots indicate the measured points.
    }
    \label{fig:6}
\end{figure}

Magnetic field dependent magnetization measurements were performed in the temperature range from 2 to 50~K in zero-field-cooled conditions by cycling the field from $+$H$_{max}$ to $-$H$_{max}$ and then back from $-$H$_{max}$ to $+$H$_{max}$. The representative $M(H)$ curves are depicted in Fig. \hyperref[fig:6]{6(a)}. For $T$ $<$ 8~K, S-shaped $M(H)$ curves are found at low fields, similar to those reported for stage-2 FeCl$_3$-GIC \cite{Suzuki1998}. As seen in Fig. \hyperref[fig:6]{6(a)}, the magnetization does not show any clear tendency towards saturation. Furthermore, when cycling the magnetic field we did not find any noticeable hysteresis effects, which is consistent with the very low values of TRM discussed in the previous section.  

At 2~K and under an applied magnetic field of 7~T, the magnetization reaches approximately 3~$\mu_B$/Fe atom. Well above the spin glass transition, at 50~K, and under a field of 7~T the magnetization still reaches a relatively large value of almost 1~$\mu_B$/Fe atom. Hohlwein et al. performed $M(H)$ measurements on stage-1 FeCl$_3$-GIC by applying fields up to 8~T and reported very similar values for the magnetization at high magnetic fields \cite{Hohlwein1974}. 

The temperature dependence of the numerical derivative $(\partial M /\partial H)_T$ is depicted in Fig. \hyperref[fig:6]{6(b)}. Below approximately 10~K, the evolution of this derivative is not monotonic with increasing magnetic field: $(\partial M /\partial H)_T$ first decreases, goes through a minimum at $\mu_0H$ $\approx$ 2.5~T, after which it increases and goes through a maximum at a characteristic field $\mu_0 H_c$. This evolution of $(\partial M /\partial H)_T$ indicates that at intermediate fields the components of the magnetization perpendicular to the applied field grow at the expense of the component parallel to the applied field. This growth stops at $\mu_0 H_c$ and much higher fields are required to gradually align the magnetic moments and reach saturation. The values of $\mu_0 H_c$ are indicated in Fig. \hyperref[fig:6]{6(b)} and their  evolution is also visible in the contour plot of $(\partial M /\partial H)_T$ versus temperature and magnetic field, shown in Fig. \hyperref[fig:6]{6(c)}. $\mu_0 H_c$ shifts towards lower fields with increasing temperature, from $\mu_0 H_c$ = 6~T to 4.5~T at 2~K and 6~K respectively. With increasing temperature, the peak at $\mu_0 H_c$ weakens and becomes indistinct above 9~K. 

\subsection{\label{sec:3f}High field temperature dependent magnetization}

\begin{figure}[hbt!]
    \centering
    \includegraphics[width = 8.5 cm]{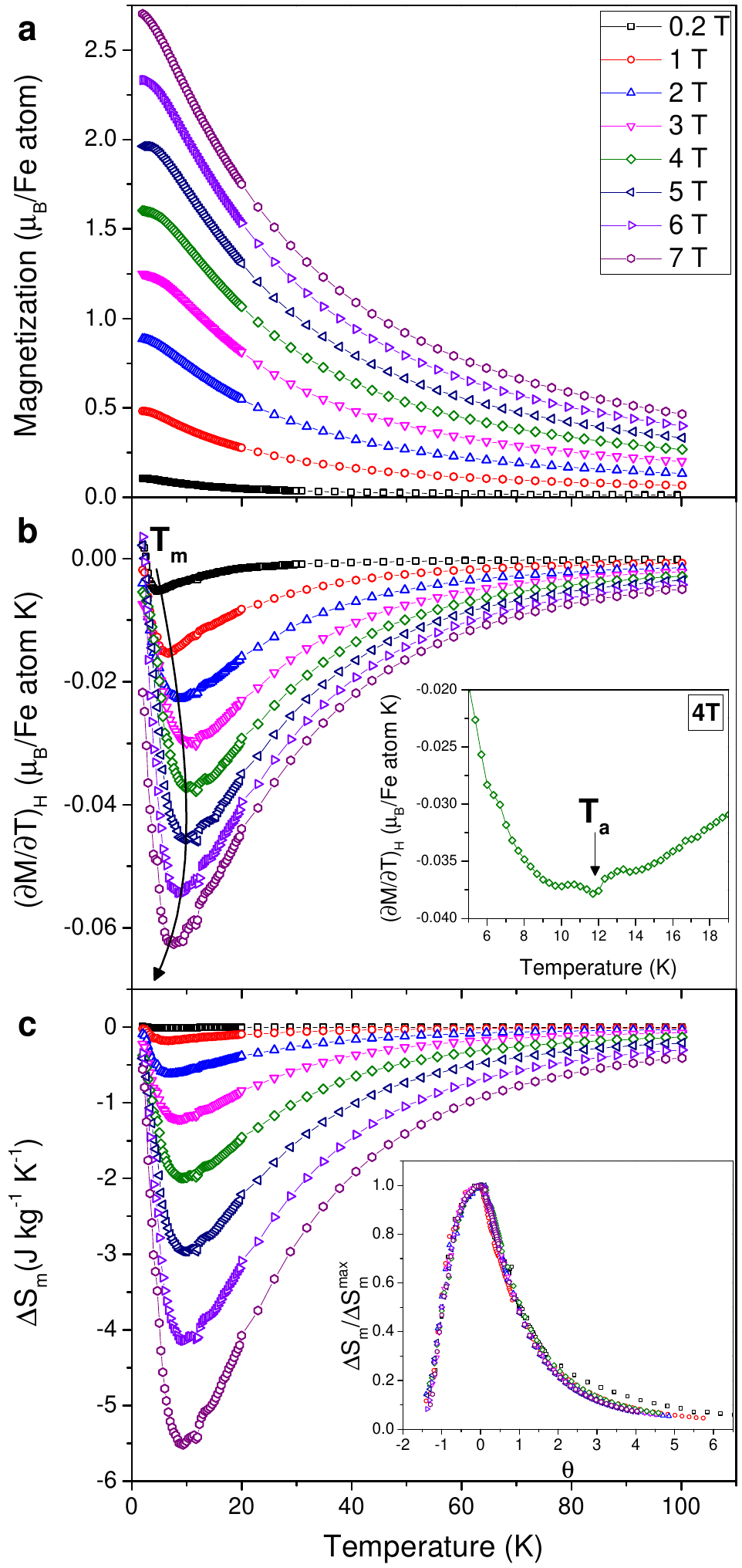}
    \caption{Temperature dependence of the magnetization (a), the derivative $(\partial M/\partial T)_H$ (b) and the deduced magnetic entropy change (c) for magnetic fields ranging from 0.2 to 7~T. The minima of the $(\partial M/\partial T)_H$ curves are indicated by $T_m$ and the shift of $T_m$ with increasing field is indicated by the black arrow. The inset of (b) shows a smaller region of temperatures, from 5 to 19~K, to highlight the observed peak at 12~K under an applied magnetic field of 4~T. The temperature, $T_a$, at which the anomaly is observed is indicated by the black arrow. The inset of (c) shows the normalized $\Delta S_M$/$\Delta S_M^{max}$ curves plotted against the reduced temperature, with $\theta$ defined by Eq. \ref{eq:9} (see text).% = ($T_c - T)/(T_{r1}-T_c)$ and $\theta_+$ = ($T - T_c)/(T_{r2}-T_c)$
    }
    \label{fig:7}
\end{figure}

In addition to the previous measurements, we also determined the magnetization as a function of temperature for applied fields varying between 1 and 7~T at intervals of 1~T. For these measurements the sample was first zero-field-cooled down to 2~K, where the magnetic field was  applied. The magnetization depicted in Fig. \hyperref[fig:7]{7(a)} was subsequently measured by increasing the temperature up to 100~K in a stepwise fashion. As shown in Fig. \hyperref[fig:7]{7}, these measurements were complemented by the magnetization data measured at 200~mT, which are discussed in section \hyperref[sec:3b]{III.B}.  

The plot of the numerical derivative $(\partial M/\partial T)_H$ against temperature, shown in Fig. \hyperref[fig:7]{7(b)}, reveals %Below 10~K, the $(\partial M/\partial T)_H$ vs. $T$ plots show 
a broad minimum at a temperature $T_M$, which first increases with increasing magnetic field up to 3~T, but then decreases for higher magnetic fields. Within this broad minimum an additional smaller anomaly can be seen at $T_a$ = 12~K, as highlighted by the inset of Fig. \hyperref[fig:7]{7(b)}. This feature does not shift with increasing field, is relatively small and seems to scale with $(\partial M/\partial T)_H$. For this reason we checked whether its origin is instrumental by measuring under the same conditions and with the same apparatus, a Co$_4$(OH)$_6$(SO$_4$)$_2$[enH$_2$] sample. This system becomes ferromagnetic below $T \approx$~13~K \cite{Levinsky2015} and at 12~K the derivative $(\partial M/\partial T)_H$ is comparable to that of our stage-1 FeCl$_3$-GIC samples. This small anomaly was absent in the data of Co$_4$(OH)$_6$(SO$_4$)$_2$[enH$_2$] but always present in the data of all our stage-1 FeCl$_3$-GIC samples, even though these came from different synthesized batches. Furthermore, the anomaly cannot be explained by any common magnetic impurities that could plausibly contaminate the samples as their ordering temperatures are significantly higher than 12~K (FeOOH polymorphs, Fe$_2$O$_3$ polymorphs, Fe$_3$O$_4$ and FeOCl) \cite{Murad1996,Grant1971}. We therefore conclude that this anomaly, the origin of which needs clarification, reflects the intrinsic magnetism of our stage-1 FeCl$_3$-GIC samples. 

As seen in Fig. \hyperref[fig:7]{7(b)}, the broad minimum in the $(\partial M/\partial T)_H$  vs. T plots reflects an inflection point of the $M(T)$ curves associated with an increase of the magnetization over a broad temperature range.
The integration of the numerical derivative  $(\partial M/\partial T)_H$ over the magnetic field leads to the change of the magnetic isothermal entropy \cite{Franco2018}: 
\begin{equation}
    \Delta S_M = \mu_0 \int_{0}^{H_{max}} \left(\frac{\partial M}{\partial T}\right)_H \, dH,
    \label{eq:7}
\end{equation}

\noindent which is plotted against temperature in Fig. \hyperref[fig:7]{7(c)}. In applied fields of 2, 5 and 7~T the maximum magnetic entropy change reaches -0.62, -2.97, and -5.52~J~kg$^{-1}$~K$^{-1}$ respectively. Such a change in entropy may be exploited for magnetic cooling applications and for this purpose a commonly used metric is the relative cooling power (RCP) \cite{Franco2012}: 

\begin{equation}
    RCP_{\text{FWHM}} = \Delta S_M^{Pk} \; \delta T_{\text{FWHM}}.
    \label{eq:8}
\end{equation}

\noindent Here $\Delta S_M^{Pk}$ is the peak value of the change in magnetic entropy on applying a magnetic field and $\delta T_{\text{FWHM}}$ is the full width at half maximum of the associated peak. For our sample the $RCP_{\text{FWHM}}$ under applied magnetic fields of 2, 5 and 7~T amounts to 14.2, 73.95 and 174.0~J/kg respectively. 

These values are significantly lower that the largest values reported so far, which are found in ferromagnetic rare-earth-based materials (see e.g. \cite{Zhang2019,Singh2007, Yi2017}), but are comparable to those found in antiferromagnetic Tm$_2$Ni$_2$Ga, Ho$_2$Ni$_2$Ga and the cluster glass  Pr$_2$Ni$_{0.95}$Si$_{2.95}$ \cite{Zhang2018,Pakhira2019}. 

The inset of Fig. \hyperref[fig:7]{7(c)} shows that the values of $\Delta S_M$ measured under various applied field strengths normalized to their maximum value, $\Delta S_M/\Delta S^{max}_M$, collapse into a single universal curve when plotted against the reduced temperature $\theta$. The latter is defined as follows \cite{Bonilla2010a}:

\begin{equation}
    \theta = \begin{cases} \theta_- = (T_c - T)/(T_{r1}-T_c),\quad T \leq T_c \\
    \theta_+ = (T - T_c)/(T_{r2}-T_c),\quad T > T_c
    \end{cases}
    ,
    \label{eq:9}
\end{equation}

\noindent  where $T_{r1}$ and $T_{r2}$ are the reference temperatures, defined such that $\Delta S_M(T_{r1},T_{r2})$ =  \( \frac{1}{2} \)$\Delta S^{max}_M$, below and above $T_c$ respectively. Such a  scaling reflects the building up of ferromagnetic correlations, and it has been argued that it indicates a second order phase transition to a ferromagnetic ground state at $T_c$ \cite{Bonilla2010a,Bonilla2010b}. However, similar behavior has also been found in systems which do not undergo a second order ferromagnetic transition, such as the chiral helimagnets MnSi \cite{Ge2015} or  YbNi$_3$Al$_9$ \cite{Wang2020}.  Also our  stage-1 FeCl$_3$-GIC sample does not become ferromagnetic for $T<T_c$, as e.g. the susceptibility does not diverge (see Fig. \hyperref[fig:7]{3}). Nonetheless, the observed behavior indicates that  ferromagnetic correlations build up with decreasing temperature with a correlation length which does not diverge but remains finite. 

\section{\label{sec:4}Discussion}
\subsection{\label{sec:4a}Origin of the magnetic phases}
As already discussed, our analysis reveals in stage-1 FeCl$_3$-GIC the existence of two magnetic phase transitions, at $T_{f1} \approx$~4~K and at $T_{f2} \approx$~3~K, below which a cluster glass-like phase and a long-range ordered phase are formed respectively. Two factors, both related to the process of intercalation, appear to be important in determining the mechanism behind these two transitions: the presence of three distinct Fe-sites at low temperatures and the formation of intercalant islands \cite{Wertheim1981,Millman1983}. 

We argue that the cluster glass-like phase below $T_{f1}$ originates from the combination of structural inhomogeneities, i.e. intercalant islands that are inherent to the intercalation process, and frustration. The latter may arise from competing interactions and anisotropies between magnetic moments on the minority Fe$^{2+}$  and the majority Fe$^{3+}$ sites \cite{Millman1983}. According to Suzuki and Suzuki \cite{Suzuki1998}, the intraplanar exchange interactions, as determined by M\"{o}ssbauer spectroscopy, should be antiferromagnetic between Fe$^{3+}$ ions and ferromagnetic between Fe$^{2+}$ ions. As the magnetic order below $T_{f1}$ should be 2D according to earlier PND results \cite{Simon1983}, the observed cluster glass-like behavior is puzzling because a 2D spin or cluster glass phase should only occur at T $=$ 0~K \cite{Binder1986}. The change of behavior at $T_{f1}$  would therefore reflect a gradual freezing of the magnetic moments, arising from large correlated volumes, instead of a phase transition. This is  consistent with the existence of ferromagnetically correlated volumes that build up above $T_{f1}$, as discussed in the previous section. This assumption would also explain the strong frequency dependence of $T_{f1}$ and the unusually low value of the exponent $z\nu$, deduced from the critical slowing down analysis discussed in section \hyperref[sec:3c]{III.C}.

The second transition temperature, $T_{f2}$, is almost frequency independent which is in line with previous PND results that indicated a transition to a long range 3D ordered phase \cite{Simon1983}. As the magnetic moments  of the  Fe$^{3+}$ ions on the (majority)  A sites lie in the basal plane \cite{Millman1983,Millman1982a}, it is plausible that the 2D magnetic correlations above $T_{f2}$ cross over to a 3D magnetic order triggered by a ferromagnetic interlayer exchange between  intercalant layers.

\subsection{\label{sec:4b}Magnetic phase diagram}

\begin{figure}[t!]
    \centering
    \includegraphics[width = 8.6 cm]{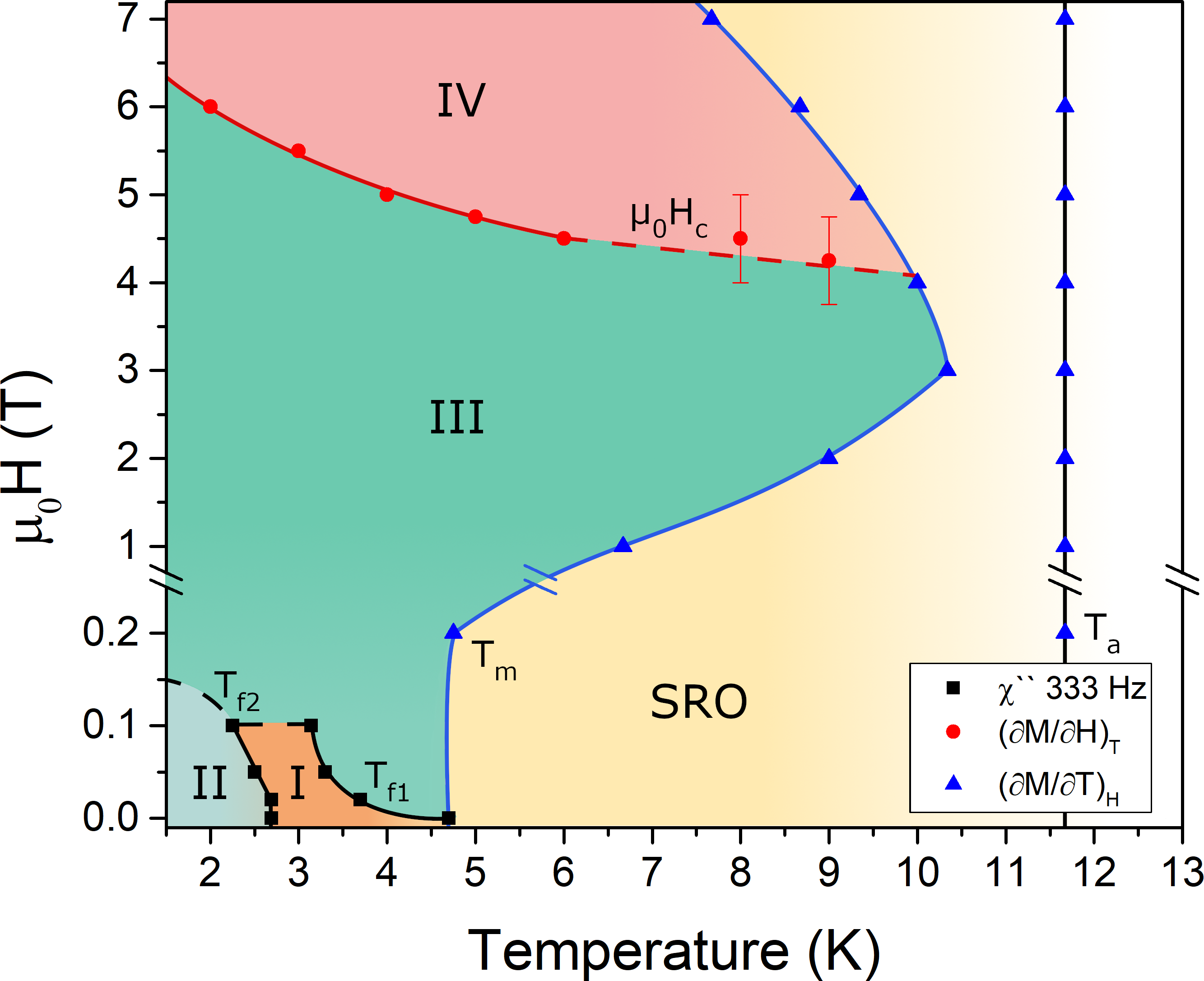}
    \caption{Magnetic phase diagram constructed from the peak positions in the AC susceptibility measurements at 333~Hz and the characteristic fields and temperatures observed in the derivatives $(\partial M/\partial H)_T$ and $(\partial M/\partial T)_H$. The characteristic field and temperature lines are labelled as $\mu_0 H_c$, $T_{f1}$, $T_{f2}$, $T_m$ and $T_a$. Part of the $T_{f1}$ and $T_{f2}$ lines are dashed to indicate uncertainty regarding up to which fields and temperatures the phases are stabilized. The $\mu_0 H_c$ line is dashed above 6~K and has large error bars reflecting the uncertainty in the determination of the center of the peaks of the derivative $(\partial M/\partial T)_H$. The short-range ordered phase is labeled as SRO and the other regions are labelled as \textbf{I} through \textbf{IV}.}
    \label{fig:8}
\end{figure}

The characteristic temperatures and fields of stage-1 FeCl$_3$-GIC identified by our  analysis are summarized in the magnetic phase diagram shown in Fig. \hyperref[fig:8]{8}.

From the AC susceptibility measurements of section \hyperref[sec:3a]{III.A} we obtain the characteristic temperatures $T_{f1}$ and $T_{f2}$.
From the $(\partial M/\partial H)_T$ curves, discussed in section \hyperref[sec:3e]{III.E}, we determine the characteristic fields $\mu_0 H_c$,  which are not very well defined above 6~K leading to large error bars.  For this reason the red $\mu_0 H_c$ line in Fig. \hyperref[fig:8]{8} is dashed above 6~K. Finally, from the $(\partial M/\partial T)_H$ curves shown in section \hyperref[sec:3f]{III.F} we obtain the characteristic temperatures $T_m$ (blue line in Fig. \hyperref[fig:8]{8}) and $T_a$. As discussed in section \hyperref[sec:3f]{III.F}, the origin of $T_a$ is unclear and further investigations are  required to characterize its nature and its effect on the magnetic correlations.

Based on the characteristic lines shown in Fig. \hyperref[fig:8]{8} we can identify several regions in the phase diagram. The cluster glass-like and long-range ordered phases, labelled as \textbf{I} and \textbf{II}, are stabilized below the $T_{f1}$ and $T_{f2}$ lines respectively. The regions labelled as \textbf{III} and \textbf{IV} extend up to $T_m$ and are separated by the $\mu_0 H_c$ line. Finally, above the $T_m$ line, the short-range correlated paramagnetic phase sets in, which is labelled as SRO in Fig. \hyperref[fig:8]{8}. 
As discussed in section \hyperref[sec:3e]{III.E}, region \textbf{III} is characterized by the growth of the components of the magnetization perpendicular to the applied field. This is reminiscent of  spin flop  or metamagnetic transitions  in antiferromagnets \cite{Liu2015a,Li2009,Tobia2008}, vortices in reentrant spin glasses \cite{Mirebeau2018}, or of the topologically non-trivial magnetic skyrmion lattice phases in chiral magnets, such as MnSi \cite{Bauer:2012cw} or Cu$_2$OSeO$_3$ \cite{Qian2016}. The stabilization of skyrmions in stage-1 FeCl$_3$-GIC remains a possibility as centrosymmetric frustrated trigonal systems are predicted to host magnetic skyrmions \cite{Leonov2015,Lin2018,Okubo2012}. Neutron scattering experiments are however needed to identify the true nature of this magnetic phase. 
 
The growth of the components of the magnetization perpendicular to the applied field stops at $\mu_0 H_c$, and above this line the derivative of the magnetization against the applied field decreases monotonically. However, our magnetization curves do not approach saturation, which implies that much higher applied fields than our maximum field of 7~T are required to reach the spin-polarized state, where all magnetic moments are fully aligned along the field.

The SRO phase extends up to much higher temperatures than shown in Fig. \hyperref[fig:8]{8}. In fact, in NPD experiments short-range correlations are seen up to 30~K \cite{Simon1983} and our temperature-dependent magnetization (Fig. \hyperref[fig:2]{2}), deviates from the high temperature Curie-Weiss behavior below 80~K. 
When approaching $T_m$ from above, the magnetization increases rapidly, leading to the minima of the $(\partial M/\partial T)_H$ curves, that define $T_m$ and reflect a large change of the magnetic entropy. Furthermore, for $\mu_0 H$ $<$ 3~T, $T_m$ shifts to higher temperatures with increasing magnetic field. Based on these observations we conclude that the SRO phase has a strong ferromagnetic component. This however would saturate for $\mu_0 H$ $>$ 3~T, leading to the subsequent decrease of $T_m$ with increasing magnetic field. Thus, the non-monotonic magnetic field dependence of $T_m$  would reflect a crossover from ferromagnetic to antiferromagnetic dominated short-range order with increasing magnetic fields.

\section{\label{sec:5}Conclusion}

In this work, we have synthesized phase-pure stage-1 graphite intercalated FeCl$_3$ and investigated its magnetic properties. DC magnetization measurements in low fields show that the dominant interactions are antiferromagnetic at high temperatures and that their strength increases with reducing temperature. 

The ZFC and FC DC magnetization curves, measured at $\mu_0H$ = 200~mT, split below $T \approx$~5~K and the ZFC curve goes through a maximum at $T_g$ = 2.5~K, a temperature that increases with decreasing magnetic field, with $T_g$ = 3.8~K for $\mu_0H$ = 3~mT.  AC magnetic susceptibility measurements confirm these findings and show that stage-1 FeCl$_3$-GIC undergoes two magnetic phase transitions at temperatures $T_{f1} = $ 4.2~K and $T_{f2} = $ 2.7~K, determined for $f$ = 13~Hz and at zero magnetic field.

Our analysis of the frequency-dependent shift of $T_{f1}$ leads  to the conclusion that this temperature marks the onset of a low temperature cluster glass-like state with slow magnetization dynamics. This magnetic cluster glass-like state could originate from the interplay  between  inhomogeneities, such as interacting magnetically ordered intercalant islands, 
and  frustration. The latter would arise from competing spin anisotropies and exchange interactions of Fe$^{2+}$ and Fe$^{3+}$ ions.  
The AC susceptibility  further reveals a frequency-independent transition at $T_{f2}$, that we attribute to a transition to a 3D long-range ordered state. Furthermore, we observe slow thermoremanent magnetization relaxation for  $T < T_{f1}$, that persists even for $T < T_{f2}$. We have attributed this effect to magnetic moments of Fe$^{2+}$ and Fe$^{3+}$ ions on the minority B and C sites respectively, which do not participate in the long range magnetic ordered phase. 

Under magnetic fields, our results  show that the components of the magnetization perpendicular to the applied magnetic field grow at the expense of the component parallel to the applied field. This process saturates at the characteristic field $\mu_0 H_c$, which at $T$ = 2~K is equal to 6~T, a value which decreases with increasing temperature.

In the paramagnetic phase, the magnetization, measured at several selected magnetic fields, increases significantly with decreasing temperature around 10~K. This reflects a change in the magnetic entropy, which we attribute to short-range correlations with ferromagnetic character that build-up when the temperature approaches $T_{f1}$ from above. 

The  magnetic phases  found in this work lead to a detailed magnetic phase diagram, which illustrates the complex  magnetic behavior of stage-1 FeCl$_3$-GIC and can serve as a reference for future investigations searching for topological non-trivial phases in this system.

\begin{acknowledgments}
We thank Ing. J. Baas for invaluable technical advice. J.J.B.L. was supported by the research program “Skyrmionics: towards new magnetic skyrmions and topological memory” of the Netherlands Organization for Scientific Research (NWO, project 16SKYR03).
\end{acknowledgments}

\end{document}

\preprint{APS/123-QED}

\title{Supplemental material for: Magnetic phase diagram and cluster glass-like properties of stage-1 graphite intercalated FeCl\texorpdfstring{$_3$}{3}}

\author{J.J.B. Levinsky}
\email{j.levinsky@rug.nl}
\affiliation{%
 Zernike Institute for Advanced Materials, University of Groningen, Nijenborgh 4, 9747 AG Groningen, The Netherlands
}%
\author{R. Scholtens}%
\affiliation{%
 Zernike Institute for Advanced Materials, University of Groningen, Nijenborgh 4, 9747 AG Groningen, The Netherlands
}%
\author{C. Pappas}
\affiliation{
 Faculty of Applied Sciences, Delft University of Technology, Mekelweg 15, 2629 JB Delft, The Netherlands
}%
\author{G.R. Blake}
\affiliation{%
 Zernike Institute for Advanced Materials, University of Groningen, Nijenborgh 4, 9747 AG Groningen, The Netherlands
}%

\maketitle
\newpage

\begin{figure*}[h]
    \centering
    \includegraphics[width = 8.6 cm]{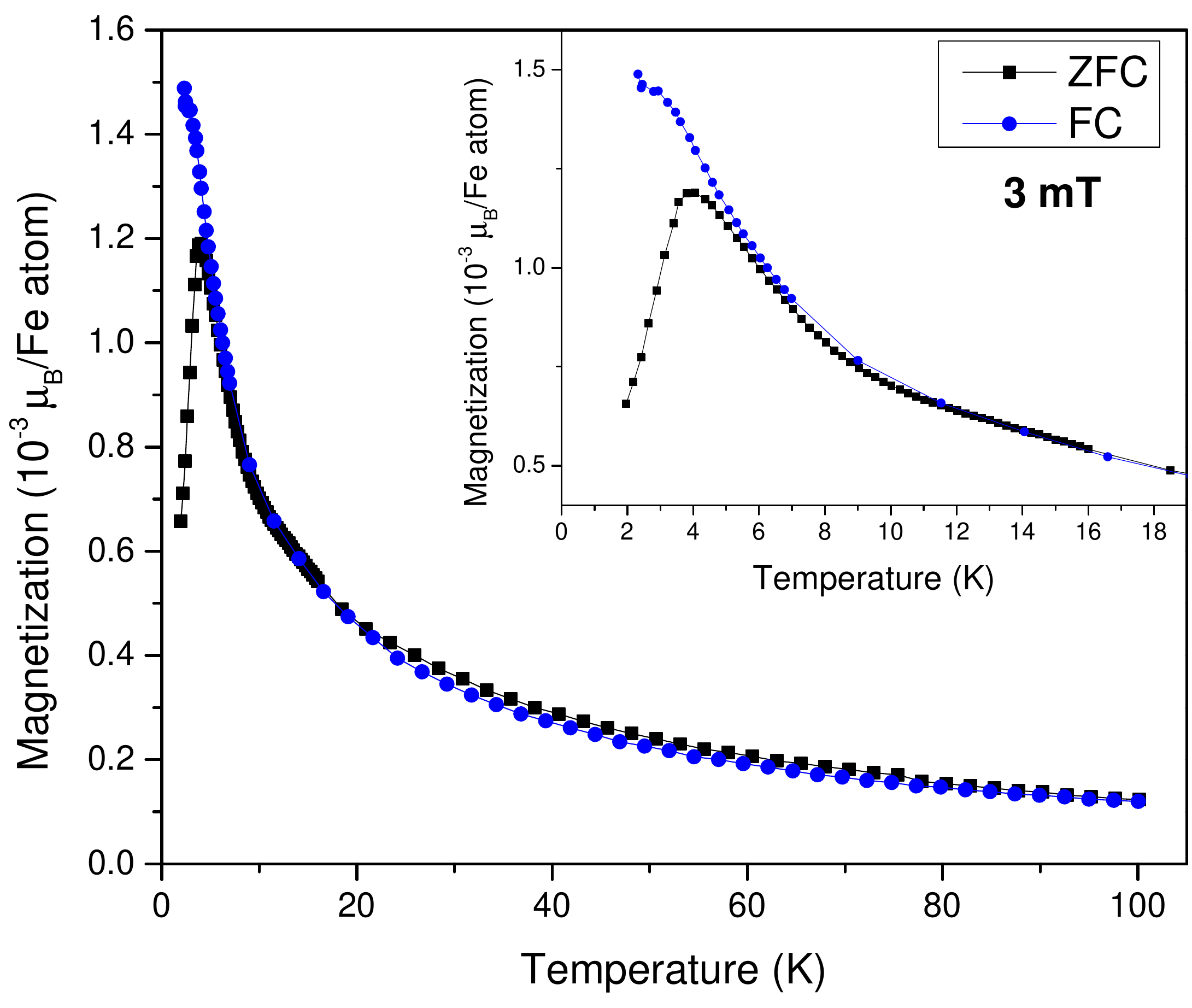}
    \caption{Field-cooled (FC) and zero-field-cooled (ZFC) magnetization versus temperature under an applied magnetic field of 3~mT. The inset shows a close-up of the low temperature region (2-19 K) of the magnetization curves to highlight the observed peak and the difference between the FC and ZFC curves.}
    \label{fig:S1}
\end{figure*}
\begin{figure*}[h]
    \centering
    \includegraphics[width = 17.2 cm]{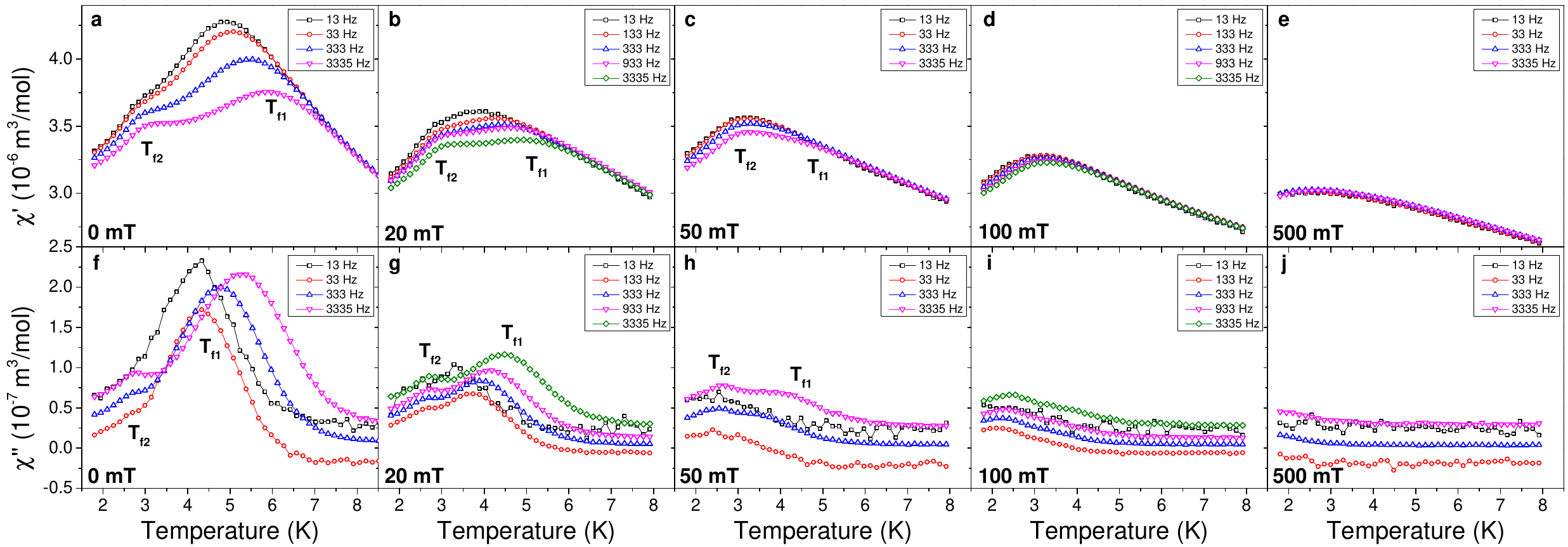}
    \caption{(a-e) Real component of the AC susceptibility plotted against temperature. The measurements were performed under a range of DC biases: 0~mT (a), 20~mT (b), 50~mT (c), 100~mT (d) and 500~mT (e). (f-j) Imaginary component of the AC susceptibility plotted against temperature. The measurements were performed under a range of DC biases: 0~mT (f), 20~mT (g), 50~mT (h), 100~mT (i) and 500~mT (j). A superimposed AC signal of 0.43~mT oscillating at the frequencies indicated in the legends was present for all measurements.}
    \label{fig:S2}
\end{figure*}